Comment on "Wavelet Analysis and scaling properties of time series"
R.B. Govindan
Graduate Institute of Technology, University of Arkansas at Little Rock, AR 72204

In a recent work Manimaran et al. [Manimaran et al., Phys. Rev. E 72, 046120 (2005)] propose to use multiresolution Daubechies (DB) wavelets to (detrend) remove the low frequency trends and subsequently to quantify the multifractal structure in a given time series. In this comment, by applying DB wavelets to the long range correlated data we show that in the presence of linear trends, the wavelets could not able to distinguish the correlations from trends. As the DB wavelets based detrending will not be able to quantify the correlations masked by trends, its multifractal extension can not always yield a correct estimate of the multifractal spectrum of the given data.

PACS number (s):89.65.Gh, 05.45.Df, 5.45.Tp

## 1. INTRODUCTION

Detrended fluctuation analysis (DFA) is one of the commonly used methods to quantify the correlations in a given data[1]. In this method, a profile function is obtained by integrating a zero mean data. The profile function is then divided in to disjoint segments of size $s$ and the data inside each window are fitted by an $m-th$ order polynomial $p^m$. To this end the fluctuation function $F(s)$ (at scale $s$) is obtained as the root mean square deviation of the profile from the best fits. In order to get the relation between $F(s)$ and $s$ the analysis is repeated for different values of $s$. For long range correlated data $F(s)$ follow a power-law[2,3,4]: $F(s) \sim s^{\alpha}$, where $\alpha$ is called the fluctuation exponent. This simple scaling relation indicates that the fluctuations at the small and large $s$ values scale to the same extent (i.e.) $s^\alpha$. This is valid only for monofractal data and is difficult to infer the signatures of multifractality from this simple scaling relation. For this purpose, a new method *mf*-DFA is proposed by considering the higher order moments of the deviations of the profile from the best fits[5].

In DFA, the trends captured by the local polynomials at scale $s$ correspond to the low pass filter of the profile with filter cut off being $sf/s$, where $sf - Hz$ is the sampling frequency of the time series ($sf = 1Hz$ for interval data). Thus, any alternative approach to remove the low frequency components of the profile can be used to replace the job of local polynomials. One of the possible approaches is attempted by Manimaran et al.[6] using Daubechies (DB) wavelets[7]. In this approach the given data of length $N$ (usually in power of 2) sampled at a frequency of $sf - Hz$ are decomposed into a set of approximate (low-pass) and detailed (high pass) coefficients corresponding to the frequency (or window size $s$) dictated by the level of decomposition. For instance, $m-th$ level of decomposition will result in $N/2^m$ number of approximate coefficients and same number of detailed coefficients with $2^m$ being the window size $s$. The approximate coefficients at the $m-th$ level is equivalent to down sampling the data into $N/2^m$ samples or $sf/2^m - Hz$ or $sf/s - Hz$. Thus, at $m-th$ level of decomposition, the reconstructed series using the approximate coefficients will correspond to the low pass filter of the profile with the filter cut off being $sf/s - Hz$. To this end the fluctuation function $F(s)$ is obtained as the mean of the magnitude of deviation of the profile from the reconstructed series[6]. To capture the multifractal behavior, higher order moments of the magnitude of the deviations of the profile from the reconstructed series are considered[6].

Though the idea of employing wavelets to capture the low frequency components of the profile is correct, here we show that in the presence of trends, it is not possible to characterize the correlations correctly using DB wavelets. As the wavelets based detrending already has problems in quantifying the correlations (monofractality) in the time series masked by trends, we conclude that it's multifractal variant as proposed in Ref. [6] will not be able to correctly capture the higher order moments of the given time series.

## II. ANALYSIS

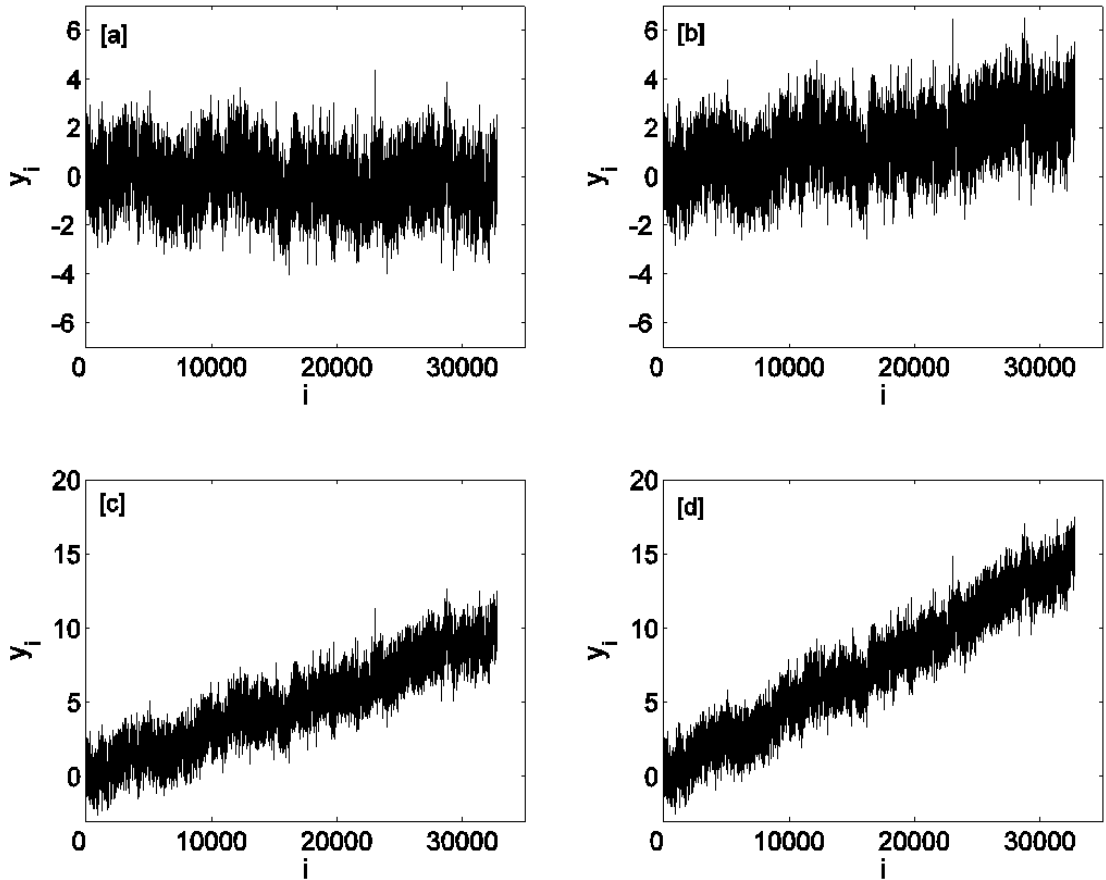

**FIG. 1.** Long range correlated data with exponent $\alpha = 0.9$ masked by different amount $c$ (see text for details) of linear trends. (a) $c = 0$ (b) $c = 3$ (c) $c = 10$ and (d) $c = 15$.

As we do not have access to the datasets used by Manimaran et al. we work with the numerically simulated long range correlated data. To emphasize the problems with wavelet based detrending approach we consider here the long range correlated data superimposed by linear trends. The three time series shown in Fig. 2 of Ref. [6] look like containing linear trend and hence we consider the effect of linear trends on the long range correlated data. The results of the raw data without any trend will serve as control in all the cases considered here.

For the present study we computed the fluctuation exponent from Fig. 4 of Manimaran et al[6] as 0.9. We generated long range correlated data $x_i$ with $\alpha = 0.9$ using Fourier filtering technique[8]. To this data we added linear trend as follows: $y_i = x_i + c \cdot i / N$. We consider $y_i$ for four different values of $c$ as shown in Fig. 1. Sampling frequency of all the data is assumed to be $1-Hz$. The slope of $y_i$ shown in Fig. 1 (b-d) increases with increase in the magnitude of $c$. It is worth to mention that DFA with second order polynomial detrending would completely remove this linear trend[9]. The fluctuation functions obtained from 8-th order DB wavelets for the four different datasets shown in Fig. 1(a-d) are given in Fig. 2. To avoid the edge effects introduced by the wavelets at longer time scales[6], we applied the wavelet based detrending from beginning to end of the profile and repeated the same from the reverse side of the profile[10] and computed the average of the fluctuations obtained in either directions. To see the effect the trends we scale the fluctuation functions by the expected scaling behavior of $s^{0.9}$ and we use this convention throughout the manuscript. If the employed wavelet correctly distinguishes the correlations from trends, the scaled fluctuation functions should be parallel to abscissa.

The scaled fluctuation function for $c=0$ is parallel to abscissa indicating that the wavelet correctly captures the correlations in the dataset. However, for large time scales, there is a slight deviation for $c=3$ ($s>10^3$) and marked deviations for $c=10$ and 15 ($s>200$) from the expected scaling behavior. Thus, the wavelets could not able to clearly distinguish the correlations from linear trends.

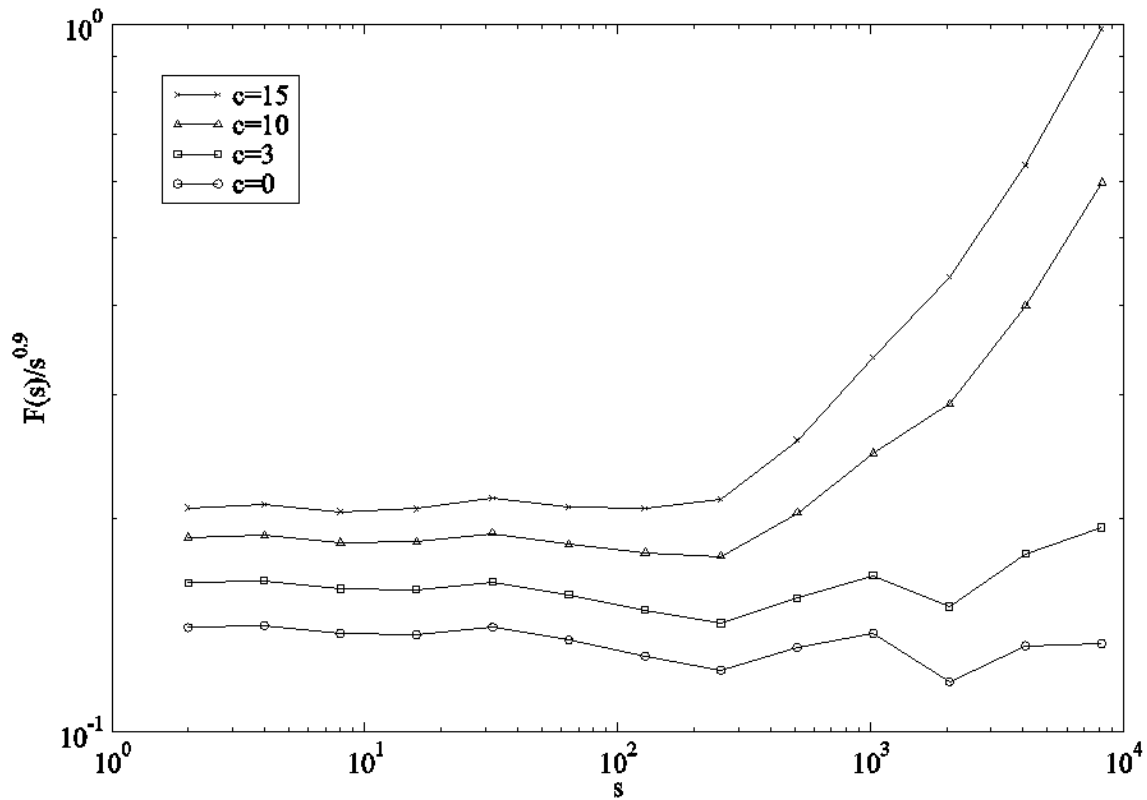

**FIG. 2** Log-Log representation of the fluctuation functions obtained using DB8 wavelets for the datasets shown in Fig. 1. The fluctuation functions are scaled by the expected scaling behavior of $s^{0.9}$.

In order to check whether the higher order wavelets could resolve this problem we computed the fluctuation functions for the same four data sets shown in Fig. 1 using 4 different higher order wavelets viz. DB16, DB24, DB32 and DB40. These results are given in Fig. 3. The results obtained using higher order wavelets are qualitatively similar to those obtained using DB8 wavelets (Fig. 2). For larger time scale $s > 300$ the deviations from the expected scaling behavior are analogous to those observed with the DB8 wavelets (see Fig. 2) though there is a slight shift in the cross-over from $s = 300$ to $s = 400$ for DB32 and DB40 (see Fig. 3 (c) and (d)). These results clearly show that wavelet based detrending might not yield correct quantification of underlying correlations when the data are masked by linear trends.

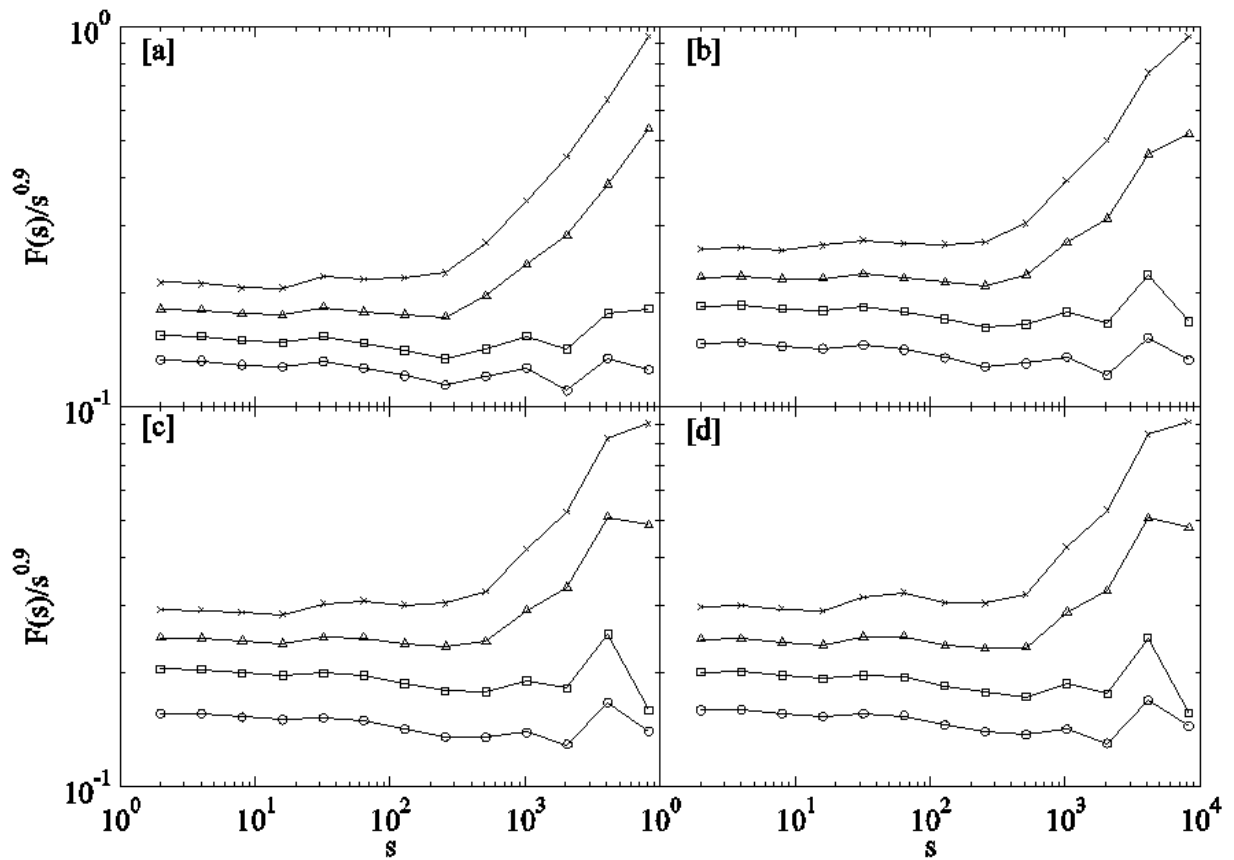

**FIG. 3** The fluctuation functions obtained using different higher orders of wavelets (a) DB16 (b) DB24 (c) DB32 and (d) DB40 for the datasets shown in Fig. 1. The fluctuation functions are scaled by the expected scaling behavior of $s^{0.9}$. Different symbols have the same meaning as in Fig. 2.

### III. DISCUSSION

In wavelet analysis, the approximate and detailed coefficients are obtained as the weighted averages (low-pass) and differences (high-pass) of the given dataset (which is profile in the present case). Reconstructing the profile from the approximate coefficients of a particular level, though expected to remove the trends of certain nature (to which the chosen order of DB wavelet is presumed to be orthogonal), the way by which the trends are removed by the wavelets is not straight forward. In the multifractal formalism, if the profile is not detrended properly, the spurious crossover in the fluctuation function due to trends will be further magnified when the higher positive moments are considered. This crossover will mislead to a slightly broader multifractal spectrum even for monofractal data with trends. For example, consider the three time series shown in Fig. 2 in Ref. [6]. All of them contain a clear linear trend. Though their magnitude is not as same as used in the present study, this type of trend, because of the problems with the wavelets in distinguishing correlations from trends, can not be removed completely by the wavelets. The variation of scaling exponents $\tau(q)$ as a function of different moment $q$ (Fig. 4 in Ref. [6]) is nonlinear indicating the presence of the multifractal signatures for the datasets considered. This nonlinear relation between $\tau(q)$ and $q$ might as well be due to the in capability of the wavelets in detrending the data. However, the same datasets, with the improved approach are revisited in Ref. [10]. One can clearly see a cross over in Fig. 4 in Ref. [10] caused by the effect of the trends. Since the same datasets were used for the both studies Ref [610], the nonlinear variation of $\tau(q)$ in Fig. 4 of Ref. [6] can be very well due to the effect of trends.

## IV. CONCLUSION

By applying different higher orders of wavelets to numerically simulated long range correlated data we have shown that wavelet based detrending might not perform well in the presence of the linear trends. In real life data one does not know *a priori* the nature of the trends present in the data and hence the approach as taken in Ref. [6] can not be reliably applied to them.